\documentclass[aps,prl,final,twocolumn,groupedaddress,
               showpacs,showkeys]{revtex4}       
               
\usepackage[latin1]{inputenc}                    
\usepackage{graphicx}                            
\usepackage{latexsym}                            
\usepackage{amsfonts}                            
\usepackage{amssymb}                             
\usepackage{amsmath}                             
\usepackage[mathscr]{eucal}                      
\usepackage{dcolumn}                             
\usepackage{theorem}                             
\usepackage{hyperref}                            

\graphicspath{{.}{graphics/}}                    
\newcommand{\imag}{\mbox{i}\ }

\hypersetup{
  debug=false,                                    
  a4paper=true,                                   
  dvipdf,                                         
  pdfpagemode={UseOutline},                       
  pdftitle={Transverse Structure of Nucleon Parton Distributions from Lattice QCD}, 
  pdfauthor={LHPC/MIT collaboration},             
  pdfsubject={Phys Rev Lett publication},         
  pdfkeywords={12.38.Gc,13.60.Fz,Generalized parton distribution,lattice matrix elements}           
  }

\theoremstyle{plain}
\theorembodyfont{\slshape}
\theoremheaderfont{\scshape}

\begin{document}

\preprint{MIT-@@@} \title[Transverse Structure of Nucleon Parton Distributions
from Lattice QCD
]{Transverse Structure of Nucleon Parton Distributions from Lattice QCD}
\author{LHPC and SESAM Collaborations \\[.2cm]
Ph.~H{\"a}gler}  \altaffiliation{current address: Department of Physics and Astronomy, Vrije Universiteit, 1081 HV Amsterdam,
The Netherlands}

\author{J.~W.~Negele  } 
\author{D.~B.~Renner}
\author{W.~Schroers}
\affiliation{Center for Theoretical Physics, Massachusetts Institute of
  Technology, Cambridge, MA 02139}
\author{Th.~Lippert}
\author{K. Schilling}
\affiliation{Department of Physics, University of Wuppertal, D-42097
  Wuppertal, Germany}

\begin{abstract}
This work presents the first calculation in lattice QCD of  three
moments of spin-averaged  and spin-polarized generalized parton distributions in
the proton. It is shown that the slope of the associated generalized form
factors decreases significantly as the moment increases, indicating that the
transverse size of the light-cone quark distribution decreases as the momentum fraction of
the struck parton increases.

 \end{abstract}



\maketitle

%
%

\section{\label{sec:Introduction}Introduction}
High energy lepton scattering reveals the quark and gluon structure of the
nucleon by measuring matrix elements of the light-cone operator
 \begin{equation}
  \label{eq:light-cone-op}
  {\cal O}_q(x) \!=\!\int \frac{d \lambda}{4 \pi} e^{i \lambda x} \bar
  q (-\frac{\lambda}{2}n)\!
  \not n {\cal P}e^{-ig \int_{-\lambda / 2}^{\lambda / 2} d \alpha \, n
    \cdot A(\alpha n)}
  q(\frac{\lambda}{2} n).
\end{equation}

The familiar quark distribution $q(x)$ specifying
the probability of finding a quark carrying a fraction $x$ of the
nucleon's momentum in the light-cone frame is measured by the
diagonal nucleon matrix element, $ \langle P |{\cal O}(x) | P
\rangle = q(x) $.  Expanding  ${\cal O}(x) $ in local operators via the operator
product expansion generates the tower of twist-two
operators,
\begin{equation}
  \label{eq:gen-loc-curr}
  {\cal O}_q^{\lbrace\mu_1\mu_2\dots\mu_n\rbrace} = \bar q
  \gamma^{\lbrace\mu_1} \imag\overleftrightarrow{D}^{\mu_2} \dots
  \imag\overleftrightarrow{D}^{\mu_n\rbrace} q\,,
\end{equation}
and the diagonal matrix element $ \langle P | {\cal
  O}_q^{\lbrace\mu_1\mu_2\dots\mu_n\rbrace} | P \rangle$ specifies the
$(n-1)^{th}$ moment of the quark distribution $\int
dx\, x^{n-1} q(x) $.

The generalized parton distributions  $ H(x, \xi, t)$ and  $ E(x, \xi, t)$  \cite{Muller:1994fv,Ji:1997ek,Radyushkin:1997ki} are
measured by off-diagonal matrix elements of the light-cone operator
\begin{equation}
\langle P' |{\cal O}(x) | P \rangle\! = \!\langle\!\langle \not n \rangle\!\rangle
H(x, \xi, t) \!+ \! \frac{i \Delta_\nu} {2 m}  \langle\!\langle \sigma^{\mu \nu} n_\mu\rangle\!\rangle
E(x, \xi, t),
\end{equation} where
$\Delta^\mu = P'^\mu - P^\mu$, $ t  = \Delta^2$, $\xi = -n \cdot \Delta /2$, $n$ is a light-cone vector, and
$\langle \!
\langle \Gamma \rangle \! \rangle = \bar U(P') \Gamma U(P)$.
Off-diagonal matrix
elements of the tower of twist-two operators
$ \langle P' | {\cal
  O}_q^{\lbrace\mu_1\mu_2\dots\mu_n\rbrace} | P \rangle$ yield moments of the
generalized parton distributions, which for $\xi$ = 0, are
\begin{eqnarray}
 \int dx x^{n-1} H(x, 0, t) & = &   A_{n 0}(t) \nonumber \\
 \int dx x^{n-1} E(x, 0, t) &=&  B_{n 0}(t),
 \label{gff1}
\end{eqnarray}
where  $ A_{n i}(t)$ and $B_{n i}(t)$ are referred to
as generalized form factors (GFFs).

Analogous expressions in which the light-cone operator $\mathcal{O}_q(x)$ and
twist-two operators contain an additional $\gamma_5$ measure the
longitudinal spin density, $\Delta q(x)$ and spin-dependent generalized
parton distributions $\tilde H(x, \xi, t) $ and $\tilde E(x, \xi, t)$ with moments $\tilde A_{n i}(t)
$ and $\tilde B_{n i}(t)$. 
In this work, we present calculations of the generalized
form factors $A_{(n=1,2,3), 0}(t)$ and $\tilde A_{(n=1,2,3), 0}(t)$ in full QCD and discuss their physical significance.
%
%

\section{\label{sec:transverse-structure}Transverse structure of parton
distribution}

In general, $ H(x, \xi, t)$ is complicated to interpret physically because it
combines features  of both parton distributions and form factors, and depends on
three kinematical variables: the momentum fraction, $x$, the longitudinal
component of the momentum transfer, $\xi$, and the total momentum transfer
squared, $t$.  In the particular case in which $\xi = 0$, however, Burkardt
\cite{Burkardt:2000za} has shown that  $ H(x, 0, t)$, as well as its spin-dependent
counterpart  $ \tilde H(x, 0, t)$, has a simple and revealing physical
interpretation.

It is useful to consider a mixed representation in which transverse coordinates
are specified in coordinate space, the longitudinal coordinate is specified in
momentum space, and one uses light-cone coordinates for the longitudinal and time
directions: $x^{\pm} = (x^0 \pm x^3)/ \sqrt{2}$, $p^{\pm} = (p^0 \pm p^3)/
\sqrt{2}$.
Using these variables, letting $x$ denote the momentum fraction and 
$\vec b_\perp$ denote the transverse displacement (or impact parameter) of the light-cone operator relative to the proton state,  one may define an impact parameter dependent parton distribution in light-cone gauge
\begin{equation}
q(x, \vec b_{\perp}) \equiv \langle P^+, \vec R_{\perp}\!\!=0, \lambda | {\cal O}_q
(x,\vec b_\perp) |P^+, \vec R_{\perp}\!\!=0, \lambda \rangle,
\end{equation}
where
\begin{equation}
{\cal  O}_q(x,\vec b_\perp)\! = \!\int \frac{d x^-}{4 \pi} e^{i x p^+ x^-} \!\bar
  q (-\frac{x^-}{2}, \vec b_\perp)
 \gamma^+
  q(\frac{x^-}{2}, \vec b_\perp).
\end{equation}

Burkardt shows that  the generalized parton
distribution $H(x, 0, t)$ is the Fourier transform of the impact
parameter dependent parton distribution, so that
	\begin{eqnarray}
	H(x, 0, -\Delta_\perp^2) \!&=&\! \int d^2 b_\perp q(x, \vec b_{\perp})\,e^{ i \vec b_\perp \cdot  \vec \Delta_\perp}, \nonumber \\
	A_{n0}( -\Delta_\perp^2)  
	\!&=&\! \int d^2 b_\perp \int dx x^{n-1} q(x, \vec b_{\perp})\,e^{ i \vec b_\perp \cdot  \vec \Delta_\perp}\!,\quad
	\end{eqnarray}
where the second form follows from Eq.~(\ref{gff1}).
Although one normally only expects a form factor to reduce to a Fourier
transform of a density in the non-relativistic limit, Ref.~\cite{Burkardt:2000za} shows
that special features of the light-cone frame also produce this simple result in
relativistic field theory.
Thus,  $H(x, 0, t)$ specifies how the transverse distribution of
quarks varies with the longitudinal momentum fraction $x$.


Physically, we expect the transverse size of the nucleon to depend
significantly on $x$. Averaging $q(x,\vec b_\perp)$ over all $x$
produces $A_{1,0}(t)$ and thus corresponds to calculating the form factor. Hence, the average  size is characterized by the transverse rms radius $  \langle r_\perp^2 \rangle^{\frac{1}{2}}  =  \langle x_1^2 + x_2^2
\rangle^{\frac{1}{2}} = \sqrt{{\frac{2}{3}}} \langle r^2 \rangle^{\frac{1}{2}}$.
From the experimental electromagnetic form factor, the transverse rms charge radius of the  proton is 0.72 fm.  As $x \to 1$, a single active parton carries
all the momentum and the spectator partons give a negligible contribution.
In this case the active parton represents the (transverse) center of momentum, and the
distribution in impact parameter reduces to a 
delta function $\delta^2(\vec b_{\perp}) $ with zero spatial extent.
Indeed, explicit light-cone wave functions \cite{Brodsky:2000xy,Diehl:2000xz} bear out
this expectation, with the result \cite{Diehl:2002he}
\begin{eqnarray*}
&&q(x,\vec b_{\perp }) =(4\pi)^{n-1}\sum\limits_{n,c}\sum\limits_{a}\int
\left[ \prod\limits_{j=1}^{n}dx_{j} d^{2}r_{\perp j}\right] 
\\
&&\times\delta \left(
1-\sum\limits_{j=1}^{n}x_{j}\right) 
 \delta ^{2}\left(
\sum\limits_{j=1}^{n}x_{j} \vec r_{\perp j}\right)  
 \delta \left( x-x_{a}\right) 
\\
&&\times \delta ^{2}\left( \vec b_{\perp
}+(1-x) \vec r_{\perp a}-\sum\limits_{j\neq a}^{n}x_{j} \vec r_{\perp j}\right)  
\\
&&\times \Psi _{n,c}^{*}(x_{1},\ldots ; \vec r_{\perp 1},\ldots
)\Psi _{n,c}(x_{1},\ldots ; \vec r_{\perp 1},\ldots ),
\end{eqnarray*}
where $a$ denotes the index of the active parton, $n$ is the number of partons in the Fock state
and the sum over $c$ represents the sum over all additional quantum numbers characterizing the Fock state. Here, one explicitly observes $  \lim_{x \to 1}  q(x,\vec b_\perp)  \varpropto \delta^2(\vec b_{\perp})$.
Since $H(x,0,t)$ is the Fourier transform of the transverse distribution,
the slope in $-t= \vec \Delta^2_{\perp} $ at the origin measures the rms transverse
radius. As a result, we expect the substantial change in transverse size with $x
$ to be reflected in an equally significant change in slope with $x$. In
particular, as $x \to 1$ the slope should approach zero. Hence, when we
calculate moments of $H(x,0,t)$, the higher the power of $x$, the more
strongly large $x$ is weighted, and the smaller the slope should become.
Therefore, this argument makes the qualitative prediction that the slope of
the generalized form factors $A_{n 0}(t)$ and $\tilde A_{n 0}(t)$
should decrease with increasing $n$, and we expect that this effect should
be strong enough to be clearly visible in lattice calculations of these form
factors.

\begin{figure}[t]
  \begin{tabular}{*{3}{c}}
    \includegraphics[width=0.40\textwidth,clip=true,angle=0]{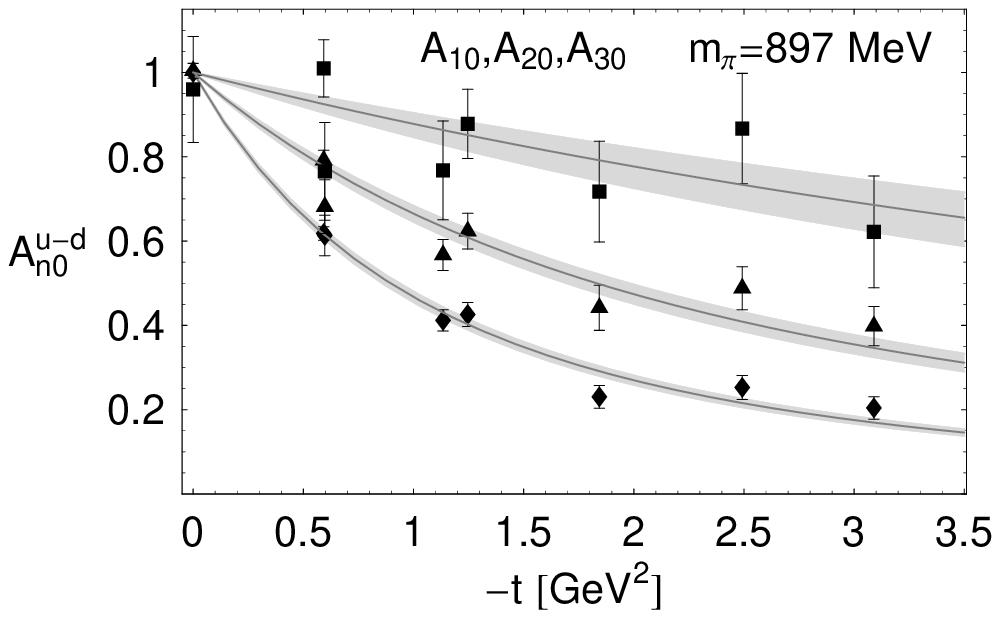}\\
    \includegraphics[width=0.40\textwidth,clip=true,angle=0]{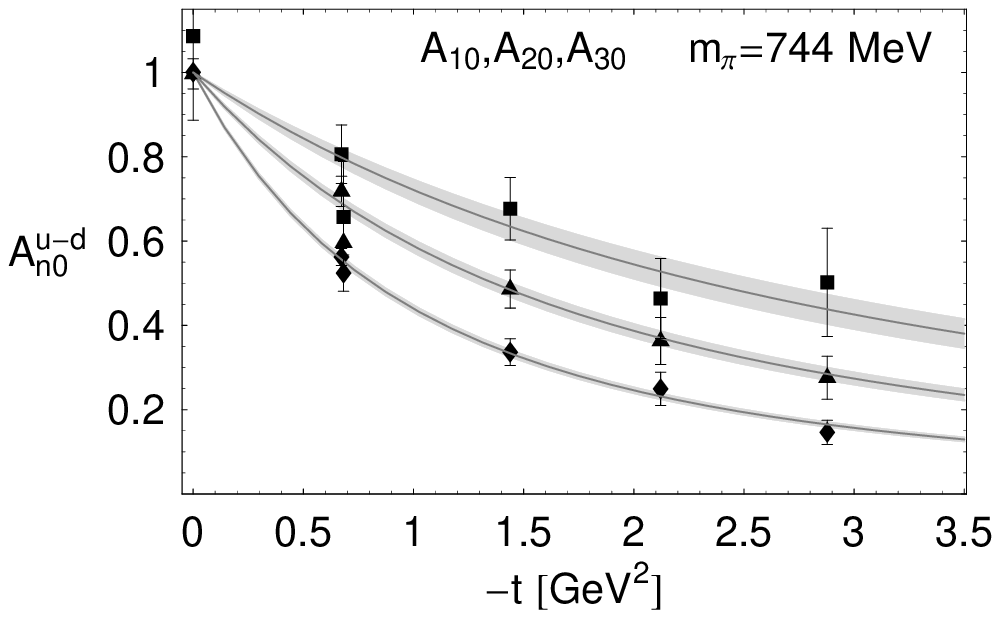}\\
    \includegraphics[width=0.40\textwidth,clip=true,angle=0]{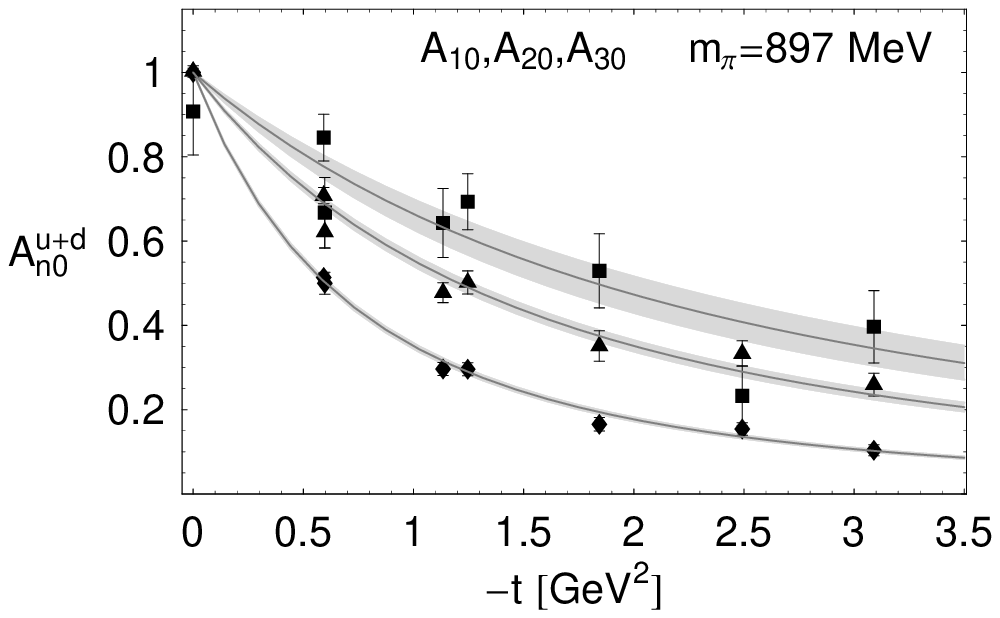}\\
    \includegraphics[width=0.40\textwidth,clip=true,angle=0]{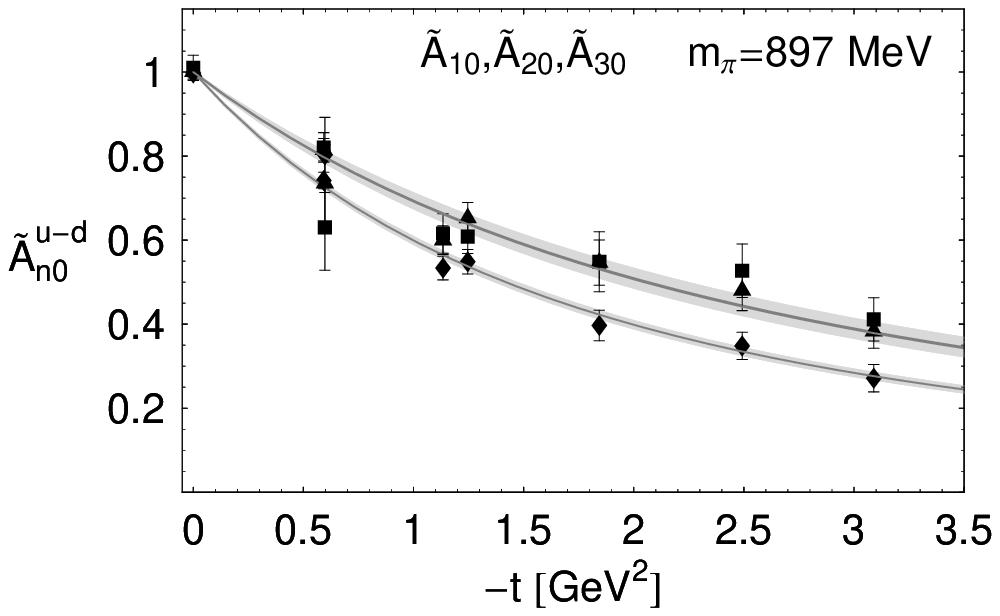}  
  \end{tabular}
  \caption{\label{fig:slopes} Normalized lattice results for generalized form factors $A_{n0}$ and  $\tilde A_{n0}$ as a function of momentum transfer squared, $-t$,
  for n=1 (circles), 2 (triangles), and 3 (squares).} 
  \label{gff_figs}
\end{figure}


%
%

\section{\label{sec:lattice-measurement}lattice measurement}

We consider the three spin independent moments 
\begin{eqnarray}
  \label{eq:explicit}
  \langle P' | {\cal O}^{\mu_1} | P \rangle  \!&=&\!
  \langle \! \langle \gamma^{\mu_1 }\rangle \! \rangle A_{10}(t)
  \nonumber \\
  &+&\! \frac{\imag}{2 m} \langle \! \langle \sigma^{\mu_1
    \alpha} \rangle \! \rangle
  \Delta_{\alpha} B_{1
    0}(t)\,, \nonumber \\ [.5cm]
  \langle P' | {\cal O}^{\lbrace \mu_1 \mu_2\rbrace} | P \rangle  \!&=&\!
  \bar P^{\lbrace\mu_1}\langle \! \langle
  \gamma^{\mu_2\rbrace}\rangle  \! \rangle
  A_{20}(t) \nonumber \\
  &+&\! \frac{\imag}{2 m} \bar P^{\lbrace\mu_1} \langle \! \langle
  \sigma^{\mu_2\rbrace\alpha}\rangle \! \rangle \Delta_{\alpha} B_{2 0}(t)
  \nonumber \\
  &+&\!\frac{1}{m}\Delta^{\{ \mu_1}   \Delta^{ \mu_2 \} }
  C_{2}(t)\,, \nonumber \\[.5cm]
  \langle P' | {\cal O}^{\lbrace\mu_1 \mu_2 \mu_3\rbrace} | P \rangle
  \!&=&\! \bar P^{\lbrace\mu_1}\bar P^{\mu_2} \langle \! \langle
  \gamma^{\mu_3\rbrace}
  \rangle \! \rangle A_{30}(t) \nonumber \\
  &+&\! \frac{\imag}{2 m} \bar P^{\lbrace \mu_1}\bar P^{\mu_2}
  \langle \! \langle \sigma^{\mu_3\rbrace\alpha} \rangle \! \rangle
  \Delta_{\alpha} B_{3 0}(t) \nonumber
  \\
  &+&\! \Delta^{\lbrace \mu_1}\Delta^{\mu_2} \langle \! \langle
  \gamma^{\mu_3\rbrace}\rangle \! \rangle A_{32}(t) \nonumber \\
  &+&\! \frac{\imag}{2 m} \Delta^{\lbrace\mu_1}\!\Delta^{\mu_2}
  \langle \! \langle \sigma^{\mu_3\rbrace\alpha}\rangle \! \rangle
  \Delta_{\alpha} B_{3 2}(t),\quad\quad 
\end{eqnarray}
where $\bar P_{\mu} = (P'_{\mu} + P_{\mu})/2$,  as well as the analogous spin dependent moments\cite{Diehl:2003ny}.



Generalized form factors $A_{(n=1,2,3),0}(t)$ and $\tilde A_{(n=1,2,3),0}(t)$ were calculated using the new method introduced in Ref.~\cite{Hagler:2003jd}. 
We considered all the combinations of $\vec P$ and $\vec P'$ that produce the same four-momentum transfer $t=(P'-P)^2 $, subject to the conditions that 
$\vec P = \frac{2\pi}{a N_s}(n_x,n_y,n_z)$ and
$\vec P'=(0,0,0)$ or $\frac{2\pi}{a N_s}(-1,0,0)$. Using all these momentum combinations  for a given $t$ below 3.5 GeV, we calculated all the hypercubic lattice operators and index combinations that produce the same continuum GFFs, obtaining an overdetermined set of equations from which we extracted a statistically accurate measurement. The errors are substantially smaller than obtained by the common practice of measuring a single operator with a single momentum combination.
Our calculations  are based on unquenched SESAM configurations \cite{Eicker:1998sy} on $16^3 \times 32$ lattices with 
$\kappa=0.1560$ and $\kappa=0.1570$,  corresponding to pion masses of $m_\pi = 897$ and  744 MeV respectively.


Figure 1 presents our principal results, showing the generalized form factors $A_{n0}(t)$ and $\tilde A_{n0}(t)$ for the lowest three moments: $n$ = 1, 2, and 3.
The form factors have been normalized to unity  at $t=0$  to make the dependence of the shape on $n$ more obvious. Note that $A_{1,0},  A_{3,0}$, and $ \tilde A_{2,0}$ depend on the difference between the quark and antiquark distributions 
whereas $ \tilde A_{1,0},  \tilde A_{3,0}$, and $  A_{2,0}$ depend on the sum.
Hence only  moments differing by two compare the same physical quantity with different weighting in $x$.
To facilitate determination of the slope of the form factors and to guide the eye, the data have been  fit using a dipole form factor
\begin{equation}
\label{dipole_fit}
A_{n0}^{\mbox{\footnotesize dipole}}=\frac{A}{\left(1-\frac{t}{m_d^2}\right)^2}.
\end{equation}
The solid line denotes the least-squares fit and the shaded error band shows
the error arising from the statistical error in the fit mass, $\Delta m_d$.  Although the dipole fit is purely phenomenological, we note that it is consistent with the lattice data. For reference, the normalization factors $A_{n0}$ and dipole masses are tabulated in Table \ref{table_values}.

The top panel in Fig.~1 shows the flavor non-singlet case $A^u - A^d$, for which the connected diagrams we have calculated yield the complete answer. It is calculated at the heaviest quark mass we have considered, corresponding to $m_\pi$ = 897 MeV. Note that the form factors are statistically very well separated, and differ dramatically for the three moments.   Indeed, the slope at the origin  decreases  by more than a factor of 2 between $n=1$ and $n=3$, indicating that the transverse size  decreases by more than a factor of 2.  The second panel shows analogous results for lighter quarks, $m_\pi$ = 744 MeV, where we observe the same qualitative behavior but slightly weaker dependence on the moment. The third panel shows the flavor singlet combination  $A^u + A^d$, for which we have had to omit the disconnected diagram because of its significantly greater computational cost. Comparing this figure with the top panel calculated at the same quark mass, we observe that while the connected contributions to $A^u \pm A^d$ are qualitatively similar, there is significant quark flavor dependence that can be used to explore the nucleon wave function. It is useful to note our results for the $u$ and $d$ GFFs are consistent with the n=2 moments calculated in Ref.~\cite{Gockeler:2003jf}.
The bottom panel shows the spin-dependent flavor non-singlet form factors $\tilde A^u - \tilde A^d$ at the heaviest quark mass. Thus, comparing the top and bottom figures displays the difference between  the spin averaged and spin dependent densities. We observe a striking difference, in that the change between the $n=1$ and $n=3$ form factors for  $q(x,\vec b_\perp)_\uparrow - q(x,\vec b_\perp)_\downarrow$ is roughly 6 times smaller than for $ \frac{1}{2} (q(x,\vec b_\perp)_\uparrow + q(x,\vec b_\perp)_\downarrow)$.   

Finally, it is useful to use the slope of the form factors at $t$ = 0 to determine the transverse rms radius,
\begin{equation}
\langle r_\perp^2 \rangle^{(n)} = {\frac{\int d^2b_\perp b^2_\perp \int dx x^{n-1} q(x,\vec b_\perp)}{\int d^2b_\perp  \int dx x^{n-1} q(x,\vec b_\perp)} }\, .
\end{equation}
Transverse rms radii calculated in this way are tabulated in Table \ref{table_values}.   To set the scale, the transverse charge radius at this mass is $ \langle r_\perp^2\rangle_{\mbox{\footnotesize charge}} $ = 0.48 fm, which is two-thirds the experimental transverse size 0.72 fm, reflecting the absence of a significant pion cloud.  For the heaviest quark mass, $m_\pi$ = 897 MeV, the non-singlet transverse size $ \langle r_\perp^2\rangle_{\mbox{\footnotesize u-d}} $ = 0.38 fm is slightly smaller than the rms charge radius, but drops 62\% to 0.14 fm for $n$=3. The singlet size $ \langle r_\perp^2\rangle_{\mbox{\footnotesize u+d}} $ is 0.46 fm, and drops 43\% to 0.27 for $n$=3.  This is a truly dramatic change in rms radius arising from changing the weighting by $x^2$. An alternative way to describe the same effect is in terms of the mean value of $x$.
The mean value of $x$ in the distribution $q(x) $ is of the order of 0.2 and roughly 0.4 in the distribution $x^2q(x)$. In these terms, the non-singlet transverse size drops 62\% as 
the mean value of $x$
increases from 0.2 to 0.4, and goes to zero when $x$ reaches 1.

\begin{table}[t]
  \begin{tabular}[b]{|c|c|c|c|}
    \hline   
     					GFF & $A(0)$ &
$m_d$ (GeV) & 		$ \langle r^2_\perp \rangle^{\frac{1}{2}}$ (fm)\\
\hline
     \multicolumn{4} {| l |}{$m_\pi$ = 897 MeV ($\kappa$ = 0.1560) } \\
\hline
     $ A_{1,0}^{u-d}(0) $     		&1.000  	$\pm$ .001
&1.470  $\pm$ .031    	& 0.380 $\pm$ 	.008  \\ \hline
     $ A_{2,0}^{u-d}(0) $     		&0.241  	$\pm$ .004
&2.102  $\pm$ .081    	& 0.266 $\pm $ 	.010  \\ \hline
     $ A_{3,0}^{u-d}(0) $     		&0.060  	$\pm$ .008
&3.857  $\pm$ .494   	& 0.145	$\pm $ 	.019  \\ \hline
     $ A_{1,0}^{u+d}(0) $     		&2.998  	$\pm$ .002
&1.205  $\pm$ .014    	& 0.463	$\pm $ 	.005 \\ \hline          
     $ A_{2,0}^{u+d}(0) $     		&0.666  	$\pm$ .009
&1.706  $\pm$ .040    	& 0.327	$\pm $ 	.008  \\ \hline
     $ A_{3,0}^{u+d}(0) $     		&0.155  	$\pm$ .018
&2.099  $\pm$ .153    	& 0.266	$\pm $ 	.019  \\ \hline
     $ \tilde A_{1,0}^{u-d}(0) $  	&1.195		$\pm$ .014
&1.850 	$\pm$ .028  	& 0.302	$\pm $  .005	 \\ \hline
     $ \tilde A_{2,0}^{u-d}(0) $  	&0.293 		$\pm$ .006
&2.223 	$\pm$ .058  	& 0.251	$\pm $  .007	 \\ \hline
     $ \tilde A_{3,0}^{u-d}(0) $  	&0.123  	$\pm$ .004
&2.233 	$\pm$ .087  	& 0.250	$\pm $  .010 	\\ \hline

     \multicolumn{4} {| l |}{ $m_\pi$ = 744 MeV ($\kappa$ = 0.1570) } \\
\hline
     $ A_{1,0}^{u-d}(0) $    	& 1.001 $\pm$ .001    	& 1.402 $\pm$ .019
& 0.398  $\pm $ .005  \\ \hline
     $ A_{2,0}^{u-d}(0) $    	& 0.261 $\pm$ .009    	& 1.814 $\pm$ .049
& 0.308  $\pm $ .008  \\ \hline
     $ A_{3,0}^{u-d}(0) $    	& 0.071 $\pm$ .013    	& 2.373 $\pm$ .138
& 0.235  $\pm $ .014  \\ \hline
    \end{tabular}
  \caption{Generalized form factors at $t=0$, dipole masses, and transverse
rms radii for the cases plotted in Fig. 1.}
  \label{table_values}
\end{table}

\section{\label{sec:summary-conclusions} Conclusions and outlook}

In the ``heavy pion world" presently accessible to full lattice QCD, we have calculated the lowest 3 generalized form factors $A_{n0}$ and $\tilde A_{n0}$ up to $|t|=$ 3 GeV as shown in Fig.~\ref{gff_figs}. We obtain excellent precision for $n$ = 1 and sufficient precision for $n$ = 2 and 3 to clearly distinguish each form factor and observe striking differences in slope and hence transverse size.  Whereas there are other calculations of isolated moments, three moments are crucial for the present investigation since $n$= 1 and 3 are necessary to measure the same combination of quark and antiquark distributions.  
The dependence of the transverse size on $x$ is most dramatic for the heaviest $u - d$ combination, for which $ \langle r_\perp^2\rangle_{\mbox{\footnotesize u-d}} $ decreases by 62\% between the first and third moment. We also observed clear dependence of the transverse distribution on flavor and spin.  Our results show that the commonly used factorization {\it Ansatz}  $H(x, 0, t) = Q(x)F(t)$ is fundamentally wrong in the ``heavy pion world" and we are aware of no arguments as to why it should be restored for lighter quarks.

 The most immediate challenges are to extend these calculations to the chiral regime of realistic quark masses, which is being explored using a hybrid calculation of dynamical staggered sea quarks and domain wall valence quarks \cite{Schroers:2003mf}, and to extend techniques for evaluating disconnected diagrams \cite{Neff:2001zr} to GFFs. When precise, controlled extrapolations to the physical pion mass are finally achieved,  moments calculated from first principles will play an essential role in complementing experimental results because of the impracticality of measuring the full $x$, $\xi$, and $t$ dependence of $H(x, \xi, t)$ and $\tilde H (x, \xi, t)$ experimentally. In addition, they will provide rich insight into the flavor and spin dependence of the transverse wave function.

%
%

\begin{acknowledgments}
P.H.~and W.S.~are grateful for
  Feodor-Lynen Fellowships from the Alexander von Humboldt Foundation
  and thank the Center for Theoretical Physics at MIT for its
  hospitality. This work is supported in part by the U.S.~Department
  of Energy (D.O.E.) under cooperative research agreement
  \#DF-FC02-94ER40818.
\end{acknowledgments}

%
%

\bibliography{GPD_References}


\end{document}